\documentclass[prl,groupedaddress,reprint,superscriptaddress,floats,nofootinbib,preprintnumbers]{revtex4-1}
\usepackage{graphicx} 
\usepackage{float}
\usepackage{amsmath}
\usepackage{epstopdf}
\usepackage{xspace}
\usepackage{rotating}
\usepackage{longtable}
\usepackage{multirow}
\usepackage{booktabs}
\usepackage[T1]{fontenc}
\usepackage{lmodern}
\usepackage[usenames,table]{xcolor}
\usepackage{comment}
\usepackage[breaklinks=true]{hyperref}
\hypersetup{
  colorlinks=true,
  linkcolor=blue,
  citecolor=blue,
  urlcolor=blue
}

\graphicspath{{./figs/}}
\usepackage{array,tabularx,epsfig,mathrsfs,rotating}
\usepackage{ifthen}
\usepackage{amsfonts}
\usepackage{amssymb}
\usepackage{slashed}

\renewcommand{\to}{\rightarrow}

\newcommand{\diag}{\text{diag}}

\newcommand{\hc}{\mathrm{h.c.}}

\newcounter{diagram}

\newcolumntype{C}[1]{>{\centering\let\newline\\\arraybackslash\hspace{0pt}}m{#1}}
\begin{document}

\begin{flushleft}
IFT-UAM-CSIC-25-43
\end{flushleft}

\begin{flushright}
COMETA-2025-17
\end{flushright}
%

\title{\boldmath $V$-Associated Production \& Vector Boson Fusion 
\\ as an LHC Signature of CP Violation
}
\author{R. Capucha}
\email[]{rscapucha@fc.ul.pt}
\affiliation{Centro de F\'{\i}sica Te\'{o}rica e Computacional,
    Faculdade de Ci\^{e}ncias
 Universidade de Lisboa, Campo Grande, Edif\'{\i}cio C8
  1749-016 Lisboa, Portugal}

\author{Á. Lozano-Onrubia}
\email[]{alvaro.lozano.onrubia@csic.es}
\affiliation{Instituto de Fisica Teorica, IFT-UAM/CSIC,
Cantoblanco, 28049, Madrid, Spain}
\affiliation{Departamento de Fisica Teorica, Universidad Autonoma de Madrid,
Cantoblanco, 28049, Madrid, Spain
}

\author{L. Merlo}
\email[]{luca.merlo@uam.es}
\affiliation{Instituto de Fisica Teorica, IFT-UAM/CSIC,
Cantoblanco, 28049, Madrid, Spain}
\affiliation{Departamento de Fisica Teorica, Universidad Autonoma de Madrid,
Cantoblanco, 28049, Madrid, Spain
}
\author{J. M. No}
\email[]{josemiguel.no@uam.es}
\affiliation{Instituto de Fisica Teorica, IFT-UAM/CSIC,
Cantoblanco, 28049, Madrid, Spain}
\affiliation{Departamento de Fisica Teorica, Universidad Autonoma de Madrid,
Cantoblanco, 28049, Madrid, Spain
}

\author{R. Santos}
\email[]{rasantos@fc.ul.pt}
\affiliation{Centro de F\'{\i}sica Te\'{o}rica e Computacional,
    Faculdade de Ci\^{e}ncias
 Universidade de Lisboa, Campo Grande, Edif\'{\i}cio C8
  1749-016 Lisboa, Portugal}
  \affiliation{ISEL -
 Instituto Superior de Engenharia de Lisboa, Instituto Polit\'ecnico de Lisboa
 1959-007 Lisboa, Portugal
}

\hfill\draft{ }

\begin{abstract}
We investigate the role of vector boson fusion (VBF) and associated production with an electroweak boson ($V$-AP) of beyond-the-Standard-Model Higgses at the LHC in probing CP violation in extended Higgs sectors. Resonant production of a new Higgs boson through $V$-AP/VBF subsequently decaying into a $Z$ boson and a 125 GeV Higgs boson $h$ would be a robust sign of CP violation, since $h$ has been measured to be (predominantly) a CP-even state. After identifying this and other sets of signatures which rely on $V$-AP/VBF to jointly uncover CP violation, we analyze the prospects to measure CP violation in this way for the complex two-Higgs-doublet-model at the High-Luminosity LHC. 

\end{abstract}

\maketitle

%

\noindent \textbf{I. Introduction.}~Charge-Parity (CP) is an approximate symmetry of Nature, only broken in the Standard Model (SM) by  mixing in the fermion sector and the presence of three families of quarks and/or leptons~\cite{Cabibbo:1963yz,Kobayashi:1973fv}. The presence of 
CP violation in the early Universe, together with baryon number violation (which occurs at high temperature in the SM~\cite{Kuzmin:1985mm} via sphaleron processes) and a departure from thermal equilibrium, is required to generate the observed baryon asymmetry of the Universe~\cite{Sakharov:1967dj,Cohen:1993nk,Trodden:1998ym,Morrissey:2012db,Konstandin:2013caa}. Yet the amount of CP violation present in the SM is well-known to be insufficient for the generation of this asymmetry at the electroweak (EW) scale~\cite{Gavela:1993ts,Gavela:1994ds,Gavela:1994dt}, such that new sources of CP violation beyond the SM are needed. 

A well-motivated scenario is that such new sources of CP violation reside in a non-minimal Higgs sector -- the Higgs sector of the SM is CP conserving by construction. LHC searches then offer a promising avenue to probe experimentally the effects of CP violation in Higgs physics (see e.g.~\cite{Ferreira:2016jea,Brehmer:2017lrt,Bernlochner:2018opw,Cirigliano:2019vfc,Englert:2019xhk,Fuchs:2020uoc,Bahl:2022yrs}), complementary to experimental searches of electric dipole moments (EDMs)~\cite{Griffith:2009zz,ACME:2018yjb,nEDM:2020crw,Roussy:2022cmp}. As opposed to EDMs, LHC searches may provide direct insight on the details of the CP-violating source.

\begin{figure}[h]
\begin{centering}
$\vcenter{\hbox{\includegraphics[width=0.21\textwidth]{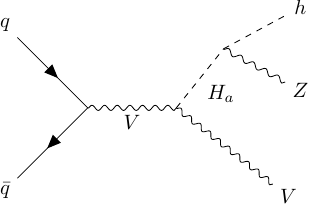}}}$
\hspace{4mm}
$\vcenter{\hbox{\includegraphics[width=0.21\textwidth]{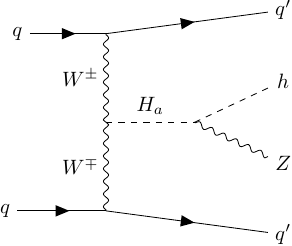}}}$
\caption{\em 
Feynman diagram for the CP-violating processes with $V$-AP production (left) and VBF production (right) with the subsequent $H_a \to Z h$ decay.}
\label{fig_Feynman1}
\end{centering}
\end{figure}

The possibility of a CP-violating Higgs sector is actively investigated at the LHC through the study of the 125 GeV Higgs boson properties, 
e.g.~via its decay $h \to \tau \tau$~\cite{Harnik:2013aja,Ge:2020mcl,CMS:2021sdq,Esmail:2024jdg} or $h \to Z Z^{*} \to 4\ell$~\cite{Bolognesi:2012mm,ATLAS:2015zhl} (with $\ell = e,\mu$) or its production in association with two jets~\cite{Dolan:2014upa} or a $t\bar{t}$ pair~\cite{ATLAS:2020ior,CMS:2020cga,CMS:2022dbt}. At the same time, LHC searches for beyond the Standard Model (BSM) scalars can yield a direct test of CP violation in extended Higgs sectors via the combination of different decay modes~\cite{Fontes:2015xva,Haber:2022gsn}, which would make inconsistent any possible assignment of CP quantum numbers for the scalars involved.
In this work we investigate a new avenue to probe CP violation within extended Higgs sectors at the LHC, by means of BSM Higgs boson  production either in association with an EW gauge boson $V = W^{\pm},\,Z$ or via vector boson fusion~\cite{Jones:1979bq} (VBF). A prime example of such a signature would be $p p \to V H_a$, $H_a \to Z h$, with $H_a$ a BSM Higgs boson (see Fig.~\ref{fig_Feynman1}-left): the $V$-associated production ($V$-AP) of $H_a$ occurs through a $g_{H_a V V}$ coupling (other contributions to $V$-AP of $H_a$ are proportional to the light fermion masses in the initial state of the partonic process, and thus negligible); the decay $H_a \to Z h$ requires a $g_{H_a Z h}$ coupling. Then, $H_a$ can neither be purely CP-even or CP-odd for both $g_{H_a V V}$ and $g_{H_a Z h}$ couplings to be present simultaneously, since the 125 GeV Higgs boson $h$ has been experimentally established as a predominantly CP-even state~\cite{CMS:2020cga,CMS:2021sdq,ATLAS:2020ior,ATLAS:2022akr,ATLAS:2023cbt}. A similar argument can be made for VBF production of $H_a$ followed by the decay $H_a \to Z h$ (see Fig.~\ref{fig_Feynman1}-right).

The above showcases that both $V$-AP and VBF production of BSM scalars $H_a$ may be key ingredients in establishing scalar CP violation at the LHC. In the following we identify sets of processes (without attempting an exhaustive classification), each set including $V$-AP and/or VBF of BSM Higgs(es), 
and for which the joint observation of all processes within a set would establish CP violation in a model-independent fashion. We then analyze the present LHC constraints on such CP-violating process combinations, together with the prospects for probing them at the High Luminosity (HL) phase of the LHC in a concrete model: the complex two-Higgs doublet model (C2HDM) with a softly-broken $\mathbb{Z}_2$-symmetry, which constitutes a minimal CP-violating Higgs sector.


\vspace{1mm}
 
\noindent \textbf{II. LHC joint signatures of CP violation.}~Considering BSM neutral Higgs bosons $H_a$ and $H_b$ (with masses $m_{H_a} > m_{H_b}$), together with the 125 GeV Higgs $h$ and the EW gauge bosons $W$ and $Z$,   
specific sets of joint LHC measurements that would yield unambiguous Higgs sector CP-violation ($H$-CPV) and include $V$-AP and/or  VBF of BSM Higgses, are:
\vspace{2mm}

\noindent \textit{(i)} $p p \to V H_a$, $H_a \to Z h$. 
\vspace{2mm}

\noindent \textit{(ii)} VBF production of $H_a$, $H_a \to Z h$.  
\vspace{2mm}

\noindent \textit{(iii)} $p p \to V H_a$, $H_a \to Z H_b$ together with $p p \to V H_b$. 
\vspace{2mm}

\noindent \textit{(iv)} VBF production of $H_a$, $H_a \to Z H_b$ together with VBF production of $H_b$.  
\vspace{2mm}

\noindent \textit{(v)} Gluon fusion (ggF) production of $H_a$, decaying via $H_a \to Z H_b$, together with both $p p \to V H_a$ and $p p \to V H_b$ (or VBF production of both $H_a$, $H_b$).  
\vspace{2mm}

\noindent \textit{(vi)} ggF production of $H_a$, decaying both via $H_a \to Z H_b$ and $H_a \to H_b H_b$ (or $H_a \to V V$), together with $p p \to V H_b$ (or VBF production of $H_b$).

\vspace{2mm}
\noindent \textit{(vii)} ggF production of $H_a$, decaying via $H_a \to Z h$, together with $p p \to V H_a$ (or VBF production of $H_a$), since $h$ is (predominantly) CP-even.
\vspace{2mm}

\begin{figure}[h]
\begin{centering}
\includegraphics[width=0.19\textwidth]{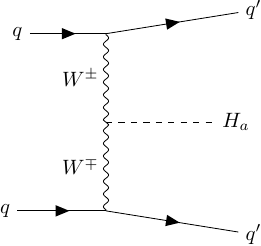} 
\hspace{6mm}
\includegraphics[width=0.19\textwidth]{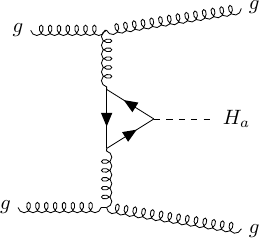} 
\caption{\em Left: Feynman diagram for VBF production of $H_a$. Right: Feynman diagram for ggF production of $H_a$ with two jets. In both cases, the final-state LHC signature is $H_a + j j$.}
\label{fig_Feynman}
\par\end{centering}
\vspace{-2mm}
\end{figure}

We view the above sets as defining potentially promising avenues to test $H$-CPV at future runs of the LHC. In all cases \textit{(i)} -- \textit{(vii)} above the $V$-AP or VBF of either $H_a$ or $H_b$ is needed to establish its coupling to EW gauge bosons ($g_{H_aVV}$ and/or $g_{H_bVV}$). For $V$-AP this is essentially granted, as the EW gauge boson can be measured as a final state of the process. This is fairly direct for leptonic decays of $V$. In contrast, for VBF the situation is much more subtle: the experimental signature of VBF is the production, e.g. of $H_a$, together with two (quark) jets -- see Fig.~\ref{fig_Feynman}-left -- with a very large di-jet invariant mass $m_{jj}$ and a large rapidity gap $\Delta \eta_{jj}$~\cite{Bjorken:1992er,Kauer:2000hi}. Yet, the ggF production of $H_a$ with two extra (gluon) jets -- see Fig.~\ref{fig_Feynman}-right -- can also produce such a $H_a + jj$ signature, yielding a potential contamination to VBF. ggF production of $H_a$ does not provide {\it a priori} information on the CP properties of $H_a$, so this ggF leakage would hinder the possibility to use VBF as a probe of $H$-CPV. 
The existence of this ggF contamination is well-known for the case of the SM Higgs boson~\cite{DelDuca:2001eu}, where a combination of cuts~\cite{Kauer:2000hi,Barger:1991ar,Rainwater:1998kj} that define VBF-enriched \textit{simplified template cross section} (STXS) regions is able to reduce the contamination of ggF into VBF to $\sim 10\%-20\%$ of the measured VBF cross section depending on the specific STXS region (see e.g. Refs.~\cite{ATLAS:2022ooq,ATLAS:2022tnm,ATLAS:2022yrq}). While these cuts could significantly reduce the ggF contamination also in BSM scenarios, the ratio of ggF to VBF production cross section of $H_a$ depends on the BSM model parameters and is {\it a priori} unknown. As a result, this prevents us from disentangling VBF from ggF in an experimental $H_a + jj$ sample, as their initial cross section ratio is not known. Thus, the ggF 
contamination constitutes a serious hurdle to the use of VBF as part of a $H$-CPV identification strategy. 

A would-be solution to this problem stems from the fact that for $p p \to H_a + j j$ production via VBF, the jets are quark-initiated, whereas for $p p \to H_a + j j$ via ggF the jets are (dominantly) gluon-initiated. Thus, efficient quark-gluon jet tagging~\cite{ATL-PHYS-PUB-2023-032,ATLAS:2023dyu} would allow to significantly reduce the ggF contamination to VBF, potentially resolving this issue. However, present quark-gluon jet tagging algorithms are not good enough for this purpose, and one would have to rely on potential future improvements of these algorithms.
For this reason, in the rest of this work we choose to leave aside VBF processes to focus on $V$-AP, and consequently on sets \textit{(i)}, \textit{(iii)}, \textit{(v)}--\textit{(vii)} above.

At the same time, we must bear in mind that the $Z$-AP of $H_a$ and/or $H_b$ may receive a (potentially dominant) contribution from ggF production of a heavier scalar (e.g. $g g \to H_a \to Z H_b$ in the case of $p p \to Z H_b$), which would mask the presence of CP violation. This prompts us to rely on $W$-AP of BSM Higgs bosons as core process within our strategy to measure $H$-CPV, for which such masking is absent.\footnote{While in extended scalar sectors there can be an $s$-channel contribution with a charged scalar $H^{\pm}$ which also contributes to $W$-AP of neutral states, such contribution is in general suppressed by the light quark masses and thus very small.} 

In the next section we analyze the possibility of probing $H$-CPV by means of 
$W$-AP, within the minimal CP-violating version of the two-Higgs doublet model: the softly-broken $\mathbb{Z}_2$-symmetric C2HDM~\cite{Ginzburg:2002wt}. The model is described in detail in App.~A. In order to avoid scalar flavour-changing neutral currents at tree level, the $\mathbb{Z}_2$ symmetry is extended to the Yukawa sector leading to four Yukawa types. We chose to work with the less constrained, the so-called Type I, where only one doublet couples to all fermions. 
Although this analysis does not exhaustively study the whole parameter space of a Type I C2HDM, it represents a proof of concept for such multi-channel searches. Moreover, notice that higher order processes induced at loop-level may in general affect the precise determination of cross sections and branching ratios. In our case, however, we expect the results illustrated in the following to hold even including these quantum effects, as the parameter space identified is very close to the alignment limit, that is the Higgs couplings are very much SM-like.

\vspace{1mm}


\noindent \textbf{III. $H$-CPV in the C2HDM.}~The C2HDM scalar potential for the two Higgs doublets $\Phi_1$ and $\Phi_2$ is given by (see e.g.~\cite{Fontes:2017zfn})
\begin{align}
\begin{split}
    V =& \,\,m_{11}^2|\Phi_1|^2 + m_{22}^2|\Phi_2|^2 - \left( m_{12}^2\, \Phi_1^{\dagger} \Phi_2 + \hc \right)+ \\
    &+\frac{\lambda_1}{2}|\Phi_1|^4 + \frac{\lambda_2}{2}|\Phi_2|^4 + \lambda_3|\Phi_1|^2|\Phi_2|^2+ \\
     & +\lambda_4(\Phi_1^{\dagger}\Phi_2)(\Phi_2^{\dagger}\Phi_1) + \left[ \frac{\lambda_5}{2} (\Phi_1^{\dagger}\Phi_2)^2 + \hc \right] \, .
     \label{eq:2HDM}
\end{split}
\end{align}
This potential implements a $\mathbb{Z}_2$ symmetry $\Phi_1 \rightarrow\Phi_1$, $\Phi_2\rightarrow-\Phi_2$, softly-broken by the $m_{12}^2$ term, and allows for both $\lambda_5$ and $m_{12}^2$ to be complex. CP violation in this model is determined by \textit{one} independent parameter, namely $\delta \equiv {\rm Arg} [(m_{12}^2)^2 \lambda_5^*]$. In the CP conserving limit, $\delta = 0, \pi$, the physical spectrum of the C2HDM has two charged scalars $H^{\pm}$, a neutral CP-odd scalar $A_0$ and two neutral CP-even scalars $H_0$ and $h_0$ (see e.g.~\cite{Branco:2011iw} for details), one of which is 
identified with the 125 GeV Higgs boson. In the presence of CP violation, $A_0$ mixes with $H_0$ and $h_0$, yielding three neutral mass eigenstates $H_1$, $H_2$, $H_3$, with $m_{H_1} \leq m_{H_2} \leq m_{H_3}$. 
The rotation angles from the original basis of neutral components of $\Phi_1$ and $\Phi_2$ to the physical basis of $H_{1,2,3}$ are denoted by $\alpha_{1,2,3}$ (with ranges $-\pi/2 < \alpha_i \leq \pi/2$). The corresponding rotation matrix $R$, parametrized in terms of $(s_i, c_i) = (\sin \alpha_i, \cos \alpha_i)$ with $i=1,2,3$~\cite{Accomando:2006asw,Khater:2003wq,ElKaffas:2007rq,Fontes:2017zfn} is given explicitly on Eq.~\eqref{rotation_matrix} of App.~A, where further details on the C2HDM can be found.

The (input) C2HDM parameters can then be chosen to be $m_{H_1}$, $m_{H_2}$, $m_{H^{\pm}}$, ${\rm Re}(m_{12}^2)$, $\alpha_1$, $\alpha_2$, $\alpha_3$, $\tan \beta$ and $v = 246$ GeV (for such choice of input parameters, the value of $m_{H_3}$ is a derived quantity, see App.~A 
for more details). 
This parameter choice is convenient, as it allows us to directly perform a large scan of C2HDM parameter points with  
\texttt{ScannerS}~\cite{Coimbra:2013qq,Ferreira:2014dya,Costa:2015llh,Muhlleitner:2016mzt,Muhlleitner:2020wwk} that satisfies present experimental constraints --  as computed via \texttt{HiggsTools}~\cite{Bahl:2022igd,Bechtle:2008jh,Bechtle:2011sb,Bechtle:2012lvg,Bechtle:2013wla,Bechtle:2015pma,Bechtle:2020pkv,Bahl:2021yhk,Bechtle:2013xfa,Stal:2013hwa,Bechtle:2014ewa,Bechtle:2020uwn} (the details on the specific implementation of experimental limits are discussed in App.~A). We pay attention to the effect of the latest EDM constraints~\cite{Roussy:2022cmp}, but do not consider them as strict bounds, as these are indirect probes of CP violation and other potential BSM sources of CP violation unrelated to $H$-CPV would also contribute to EDMs. This also allows to better explore the complementarity between EDMs and LHC probes of CP violation.

We perform two different scans of the C2HDM parameter space -- labeled $S_1$ and $S_2$ -- with the $125$~GeV Higgs boson $h$ corresponding respectively to the mass eigenstate $H_1$ and $H_2$. 
We have chosen the following regions for the input parameters of the C2HDM: $0.8 < \tan\beta < 35$, $-\pi/2 < \alpha_{1,2,3} < \pi/2$, 0.1 GeV$^2 < \text{Re}(m_{12}^2) <$ 10$^5$ GeV$^2$, 70 GeV $< m_{H^\pm} <$ 500 GeV, and for $S_1$ $m_{H_1} = 125.09$ GeV, 130 GeV $< m_{H_2} < 500$ GeV, while for $S_2$ we chose 40 GeV $< m_{H_1} < 120$ GeV, $m_{H_2} = 125.09$ GeV. Besides the use of \texttt{HiggsTools}, we further filter our scan points by requiring that they satisfy $2\sigma$ bounds from the CMS analysis~\cite{CMS:2022dwd} on the gauge and fermion coupling modifiers $c_{h V V} \equiv g_{h VV}/g_{h_{\rm SM} V V}$ and  $c_{h f f} \equiv y_{h f f}/y_{h_{\rm SM} f f}$. Concretely, we define $y_{hff} = \sqrt{(y_{hff}^R)^2 + (y_{hff}^I)^2}$ (with $y_{hff}^R$ and $y_{hff}^I$ respectively the scalar and pseudoscalar parts of the 125 GeV Higgs boson Yukawa couplings to SM fermions), and demand that the scan points lie within the $(c_{hff},c_{hVV})$ $2\sigma$ ellipse in Ref.~\cite{CMS:2022dwd} from a combined Higgs signal strength fit with observed best-fit value $(c_{hff},c_{hVV}) = (0.906,1.014)$. We emphasize that the results of this latter fit only apply to a 2HDM of Type I (addressed in this work), as all SM fermions only couple to one doublet and thus the Yukawa coupling modifiers are universal.

\begin{figure}[h]
\begin{centering}
\includegraphics[width=0.49\textwidth]{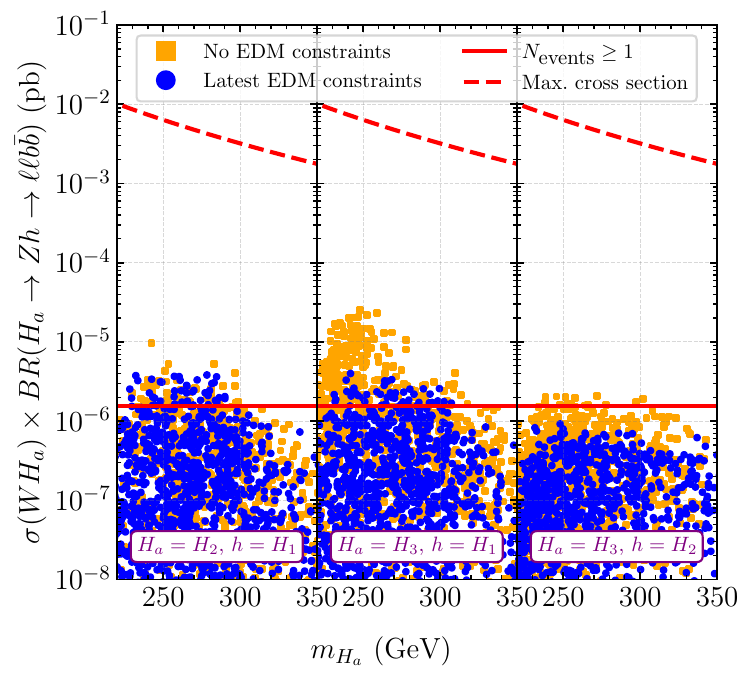} 

\vspace{-3mm}

\caption{\em $W$-AP cross sections (in pb) for $H_a$ -- with $H_a \to Zh \to \ell \ell b \bar{b}$ -- at $\sqrt{s}=14$ TeV for the C2HDM scans $S_1$ (left and central panels) and $S_2$ (right panel), as a function of $m_{H_a}$. Left panel: $H_a=H_2$. Central and right panels: $H_a=H_3$. We also depict the theoretical upper bound on the cross section (dashed-red line, see text for details) and the cross section  yielding 1 event (including the decay of the $W$-boson) at the HL-LHC (solid-red line).
}
\label{fig:Wh3_h3toh2Z}
\par\end{centering}
\end{figure}

Throughout this work, the $W$-AP cross sections of the BSM scalars are calculated at leading order for $\sqrt{s} = 14$~TeV at the LHC via \texttt{MadGraph5\_aMC@NLO 2.9.14}~\cite{Alwall:2014hca}, whereas the ggF cross sections (at $\sqrt{s} = 14$~TeV) and the branching fractions of the BSM scalars are obtained from \texttt{ScannerS} (see~\cite{Muhlleitner:2020wwk} for further details).

\vspace{1mm}

We show the cross sections for $ p p \to W H_a$, $H_a \rightarrow Z h \rightarrow \ell \ell b \bar{b}$ as a function of $m_{H_a}$ in Fig.~\ref{fig:Wh3_h3toh2Z}, with $H_a = H_2$ (left) and $H_a = H_3$ (central) for scan $S_1$, and $H_a = H_3$ for scan $S_2$ 
(right). This process corresponds to avenue \textit{(i)} to establish $H$-CPV. The theoretical upper bound on this cross section (given by $c_{H_a W W} \equiv g_{H_a WW}/g_{h_{\rm SM} W W} = 1$ and $BR(H_a \rightarrow Z h) = 1$) is shown as a dashed-red line in Fig.~\ref{fig:Wh3_h3toh2Z}. It is immediately apparent that the cross sections for this process in the C2HDM with a softly-broken $\mathbb{Z}_2$-symmetry are very suppressed, and impossible to probe at the HL-LHC with 3 ab$^{-1}$ of integrated luminosity.\footnote{Considering the leptonic decay $W \to \ell \nu$, we expect at most $\mathcal{O}(1)$ signal events for the points of scan $S_2$ at the HL-LHC, and $\mathcal{O}(10)$ signal events for the points of scan $S_1$, 
with perfect (unrealistic) final-state reconstruction efficiencies in both cases.} We leave a study of this signature in other models -- e.g. the C2HDM without the (softly-broken) $\mathbb{Z}_2$-symmetry -- for future investigation.


\begin{figure}[h]
\begin{centering}
\includegraphics[width=0.49\textwidth]{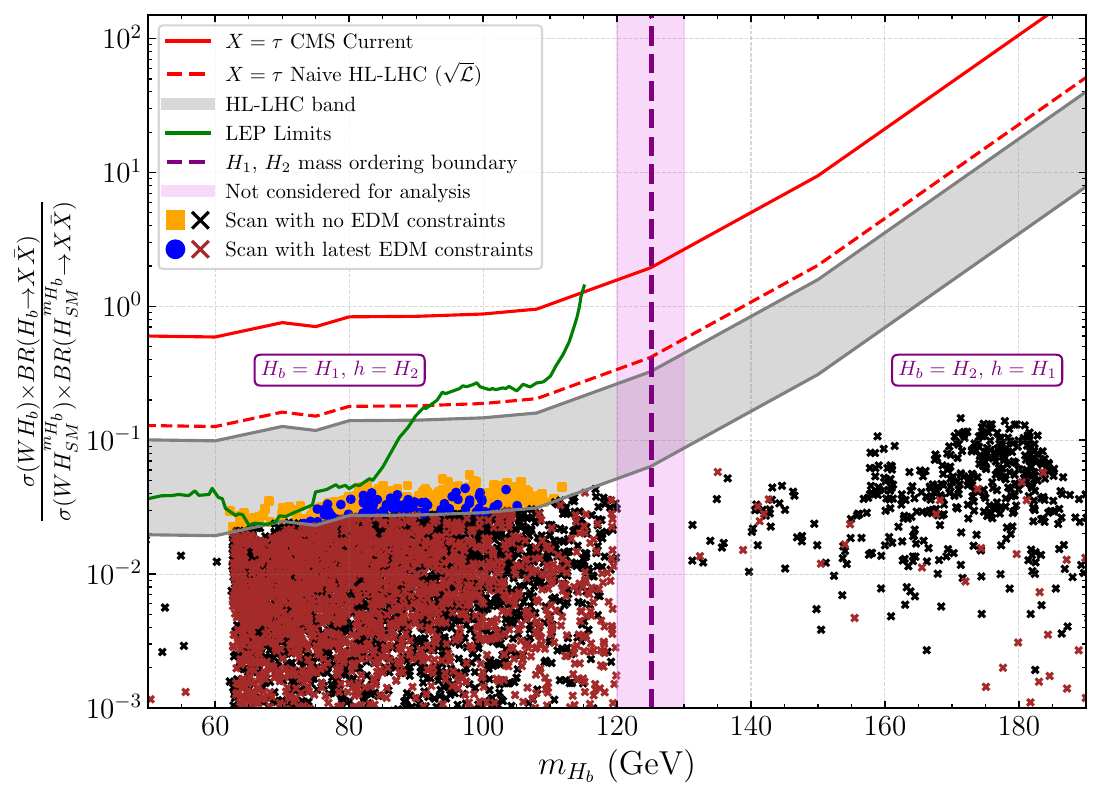} 

\vspace{-3mm}

\caption{\em Current (solid red) and projected HL-LHC (dashed-red, grey band) sensitivity lines of $pp \to W H_b$ production at $\sqrt{s} = 14$~TeV normalized to the SM associated production $\sigma^{\rm W-AP}_{SM}$ of a SM Higgs of mass $m_{H_b}$ as a function of $m_{H_b}$, see text for details. Orange squares (black crosses) correspond to HL-LHC observable (unobservable) C2HDM scan points when no EDM constraints are applied, and blue circles (red crosses) when EDM bounds are strictly imposed, respectively for scan $S_1$ (for $m_{H_b} > 125$ GeV) and $S_2$ (for $m_{H_b} < 125$ GeV). Also shown (solid green) are the LEP limits on $\sigma(W H_b)/\sigma(W H_{\rm SM}^{m_{H_b}})$ assuming $c_{H_bZZ} = c_{H_bWW}$. 
}
\label{fig:AssociatedW}
\par\end{centering}
\end{figure}

The remaining $H$-CPV sets -- except for \textit{(vii)} -- in our analysis require the measurement of $W$-associated production of a light BSM Higgs boson $H_b$, with $H_b = H_2$ ($H_b = H_1$) for scan $S_1$ ($S_2$), respectively. There exist current experimental limits on this process 
by CMS~\cite{CMS:2024ulc} for $\sqrt{s} = 13$ TeV and an integrated luminosity $\mathcal{L} = 138$ fb$^{-1}$, targeting a $H_b \to \tau \tau$ final state. These limits, normalized to the SM production cross section $\sigma^{\rm W-AP}_{\rm SM} = \sigma(pp \to W H_{\rm SM}^{m_{H_b}}, H_{\rm SM}^{m_{H_b}} \to \tau\tau)$ with $H_{\rm SM}^{m_{H_b}}$ a SM Higgs boson of mass $m_{H_b}$, are shown in Fig.~\ref{fig:AssociatedW}. To estimate the sensitivity achievable at the HL-LHC for $p p \to W H_b$, we first perform a naive ``statistical'' $\sqrt{\mathcal{L}}$ extrapolation of the current expected sensitivity of the CMS search~\cite{CMS:2024ulc} by $\sqrt{138/3000}$, also depicted in Fig.~\ref{fig:AssociatedW}. Yet, such naive $\sqrt{\mathcal{L}}$ sensitivity scaling has been found to be rather pessimistic in many existing experimental ATLAS and CMS analyses, as discussed in Ref.~\cite{Belvedere:2024wzg}. Among other things, the $\sqrt{\mathcal{L}}$ scaling does not account for the possible inclusion of further decay modes as $\mathcal{L}$ increases, and does not exploit kinematic regions (e.g. high-$p_T$ configurations) that do not appreciably contribute to the present sensitivity, but would very much do for larger $\mathcal{L}$. We take inspiration from the experimental sensitivity increase observed between $\mathcal{L} \simeq 36$ fb$^{-1}$ and $\simeq 140$ fb$^{-1}$ in various ATLAS and CMS searches performed for $13$ TeV LHC  -- specifically, $p p \to t\bar{t} t\bar{t}$~\cite{CMS:2017ocm,CMS:2023ftu}, $p p \to V h$ with $h\to b\bar{b}$~\cite{ATLAS:2017cen,ATLAS:2020fcp}, EW vector boson scattering $p p \to Z Z + j j$, $ Z Z \to \ell\ell \ell'\ell'$ (with $\ell,\ell' = e,\mu$)~\cite{CMS:2017zmo,CMS:2020fqz} and di-Higgs production $p p \to h h \to b\bar{b} \tau\tau$~\cite{ATLAS:2018uni,ATLAS:2024pov}. In all these searches, the sensitivity increase exceeds the naive $\sqrt{\mathcal{L}}$ extrapolation, albeit by different amounts. We then develop an ``improved'' HL-LHC sensitivity extrapolation band as the envelope of the sensitivity increases of the respective ATLAS/CMS analyses above, extrapolated to $\mathcal{L} = 3000$ fb$^{-1}$. Further details of this improved sensitivity extrapolation are given in App.~B. This band is shown in grey in Fig.~\ref{fig:AssociatedW}. The \textit{strongest} HL-LHC sensitivity we use in the following discussion corresponds to the lower edge of such HL-LHC extrapolation band.

We then show in Fig.~\ref{fig:AssociatedW} the results of the C2HDM scans $S_1$ and $S_2$, depicting the value of the 
$p p \to W H_b$ production cross section in the final state
$H_b\to \tau\tau$, normalized to the SM value (with $m_{h_{\rm SM}} = m_{H_b}$), and excluding the $\pm\, 5$ GeV mass window around the observed 125 GeV Higgs boson.~This is a key search for the $H$-CPV scenarios \textit{(iii)}, \textit{(v)} and \textit{(vi)}. We see that for scan $S_2$ (with $H_b = H_1$), a sizable 
amount of C2HDM points (in orange) lie above the lower-end of 
the grey band, thus representing potentially observable signals at the HL-LHC, including parameter space points that 
satisfy the latest EDM constraints~\cite{Roussy:2022cmp} (in blue). At the same time, for the scan $S_1$ there are no potentially observable points (with $H_b = H_2$) at the HL-LHC via this signature.  
We also depict in Fig.~\ref{fig:AssociatedW} the indirect limits from LEP~\cite{LEPWorkingGroupforHiggsbosonsearches:2003ing} (green line) which assume $c_{H_b ZZ} = c_{H_b WW}$ and $BR(H_b \rightarrow X\bar{X}) = BR(h_{\rm SM} \rightarrow X\bar{X})$, the latter computed for $m_{h_{\rm SM}} = m_{H_b} < 125$ GeV. These limits show the approximate suppression required for the $H_b$-gauge coupling at low masses. We also note that for $m_{H_b} < m_h/2 \simeq 62$ GeV, there are very few experimentally allowed C2HDM scan points (and no HL-LHC observable points via $p p \to W H_b$), since the stringent limits from LHC Higgs signal strengths on exotic Higgs branching fractions~\cite{ATLAS:2022vkf,CMS:2022dwd} -- here $BR(h \to H_b H_b)$ -- apply.
\begin{figure}[htbp]
\begin{centering}
\includegraphics[width=0.49\textwidth]{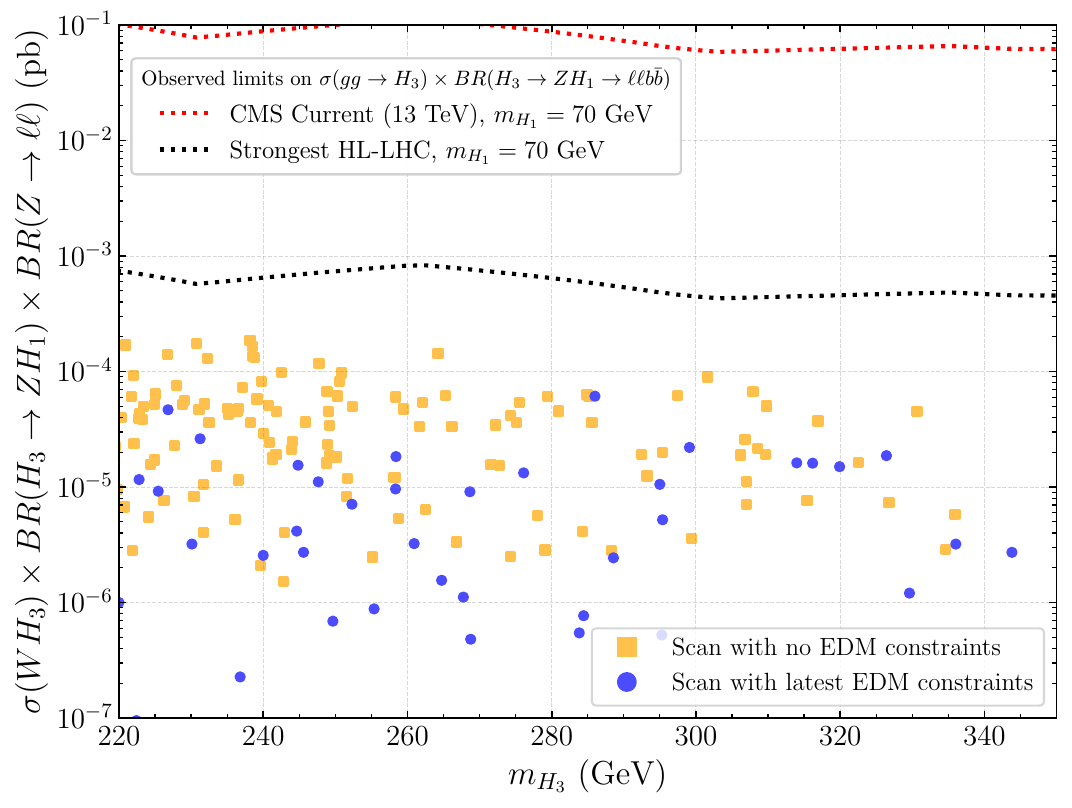}
\caption{\em $p p \to W H_3$ cross section times branching fraction BR$(H_3 \to Z H_1) \times$BR$(Z \to \ell\ell)$ for $S_2$ scan points within the HL-LHC grey band of Fig.~\ref{fig:AssociatedW}. In dotted-red (dotted-black) we show the current CMS (HL-LHC strongest projection) limits from $p p \to H_3 \to Z H_1 \to \ell\ell b\bar{b}$~\cite{CMS:2019ogx}, see text for details.
}
\label{fig:Vh3}
\end{centering}
\end{figure}

For the potentially observable C2HDM points from scan $S_2$ in Fig.~\ref{fig:AssociatedW} (in orange/blue), 
we show the cross section for $W$-AP of $H_3$ followed by its on-shell decay $H_3 \to Z H_1$ (with $Z \to \ell\ell$) in Fig.~\ref{fig:Vh3}. Observing these points would establish $H$-CPV through set \textit{(iii)}. While, to the best of our knowledge, there are no existing dedicated experimental LHC searches for this process, searches for $p p \to H_3 \to Z H_1$ by ATLAS and CMS do exist (see e.g.~\cite{CMS:2016xnc,ATLAS:2018oht,CMS:2019ogx,ATLAS:2020gxx}), focusing on ggF production of $H_3$.~Present CMS limits on $g g \to H_3 \to Z H_1 \to \ell\ell b\bar{b}$ (assuming $BR(H_1 \to b\bar{b}) = 1$) with an integrated luminosity of 35.9 fb$^{-1}$ from Ref.~\cite{CMS:2019ogx} and the potential extrapolated HL-LHC strongest reach (for a reference value $m_{H_1} = 70$ GeV) are also depicted in Fig.~\ref{fig:Vh3}, showing that even the largest C2HDM cross sections in Fig.~\ref{fig:Vh3} -- of $\mathcal{O}(0.1)$ fb -- lie well below the depicted HL-LHC reach. Despite the fact that these limits/projections do not fully apply to our case (as $W$-AP is a different $H_3$ production mode\footnote{As a result, the corresponding signal acceptances would differ from those of Ref.~\cite{CMS:2019ogx}. Additionally, the extra $W$-boson in the final state would provide new signal discrimination opportunities, which can hopefully enhance the sensitivity of the search.}) this indicates that $H$-CPV via set \textit{(iii)} will be challenging to probe at the HL-LHC.

\begin{figure}[t]
\begin{centering}
\includegraphics[width=0.49\textwidth]{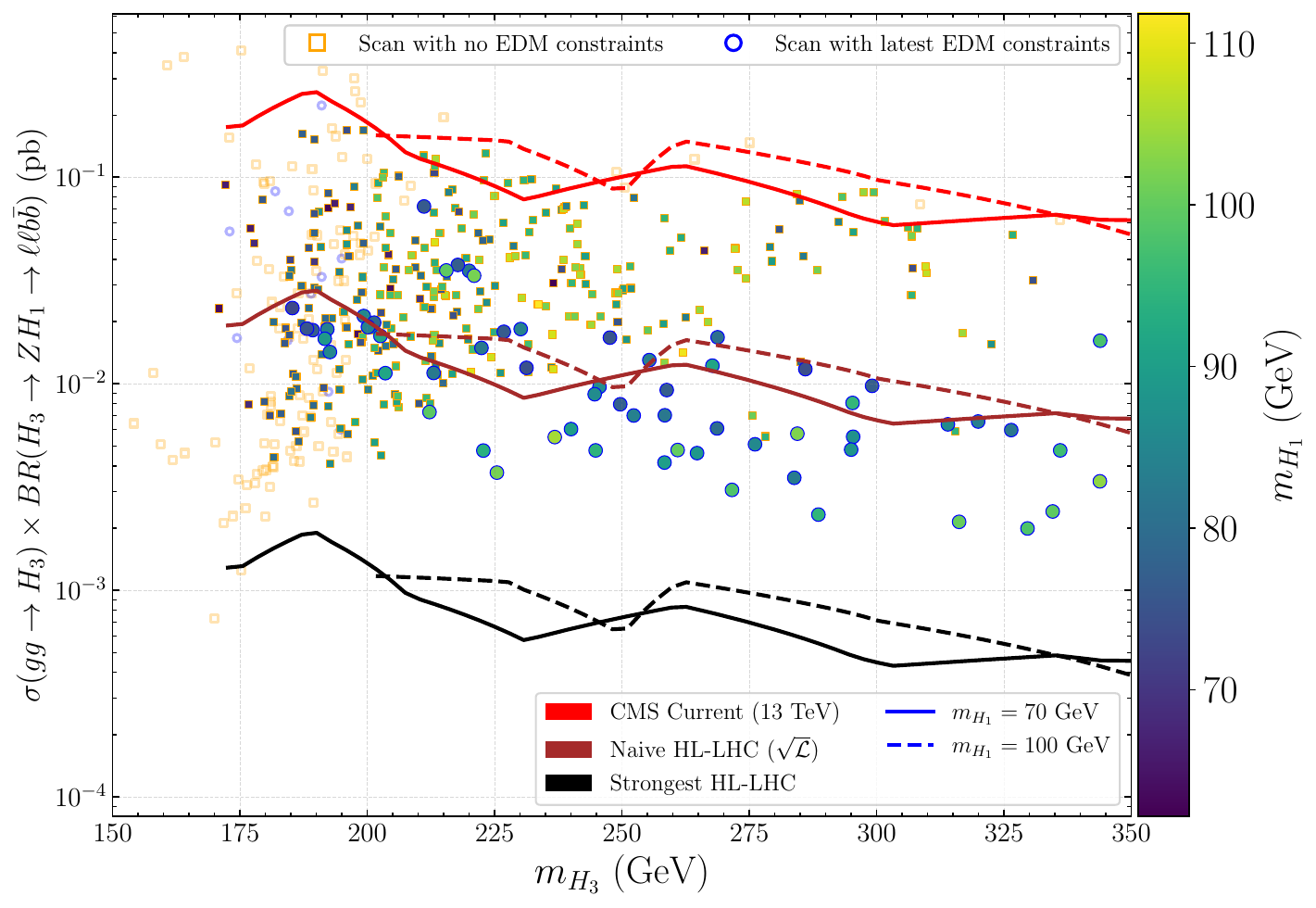}

\vspace{-3mm}

\caption{\em Cross sections for $g g \to H_3 \rightarrow Z H_1 \rightarrow \ell \ell\, b \bar{b}$ for the $S_2$ scan points within the HL-LHC grey band in Fig.~\ref{fig:AssociatedW}, as a function of $m_{H_3}$.
Squares (circles) correspond to scan points without (with) the latest EDM constraints applied. Present $\sqrt{s}=13$ TeV CMS limits from Ref.~\cite{CMS:2019ogx} for benchmark values $m_{H_1} = 70$ GeV (solid) and $m_{H_1} = 100$ GeV (dashed) are shown in red. Sensitivity projections for HL-LHC with $3000 \, \textrm{fb}^{-1}$ of integrated luminosity are shown assuming a naive $\sqrt{\mathcal{L}}$ extrapolation of the present CMS limits (brown) as well as a strongest extrapolation (black), see App.~B for details. 
Filled scan points are within reach of HL-LHC (and not excluded by present CMS constraints), with filling color indicating the corresponding value of $m_{H_1}$.
}
\label{fig:GGFH3ZH1}
\par\end{centering}
\end{figure}

\vspace{1mm}

In fact, the aforementioned signature $g g \to H_3 \to Z H_1$ is a required ingredient to establish $H$-CPV within sets \textit{(v)} and \textit{(vi)}. For the orange and blue points of Fig.~\ref{fig:AssociatedW} from scan $S_2$ that fulfill $m_{H_3} > m_{H_1} + m_Z$, we show in Fig.~\ref{fig:GGFH3ZH1} the corresponding C2HDM cross section for ggF production of $H_3$ in the $H_3 \to Z H_1$, $Z\to \ell\ell$, $H_1 \to b\bar{b}$ final state.  We overlay, for two reference mass values $m_{H_1} = 70$ GeV, $100$ GeV, the present $\sqrt{s}=13$ TeV limits (red) from the CMS search~\cite{CMS:2019ogx} (obtained from the data of Ref.~\cite{hepdata_90710}, with an integrated luminosity of 35.9 fb$^{-1}$) and the extrapolated HL-LHC reach both in the $\sqrt{\mathcal{L}}$ increase case (brown) and in the strongest case (black) as already employed in Fig.~\ref{fig:AssociatedW} and derived in App.~B.
Fig.~\ref{fig:GGFH3ZH1} clearly shows that a large fraction of the $S_2$ scan points potentially accessible via $p p \to W H_b$ searches -- recall Fig.~\ref{fig:AssociatedW} -- may also be within reach of future $g g \to H_3 \to Z H_1$ LHC searches.  

\begin{figure}[h]
\begin{centering}
\includegraphics[width=0.49\textwidth]{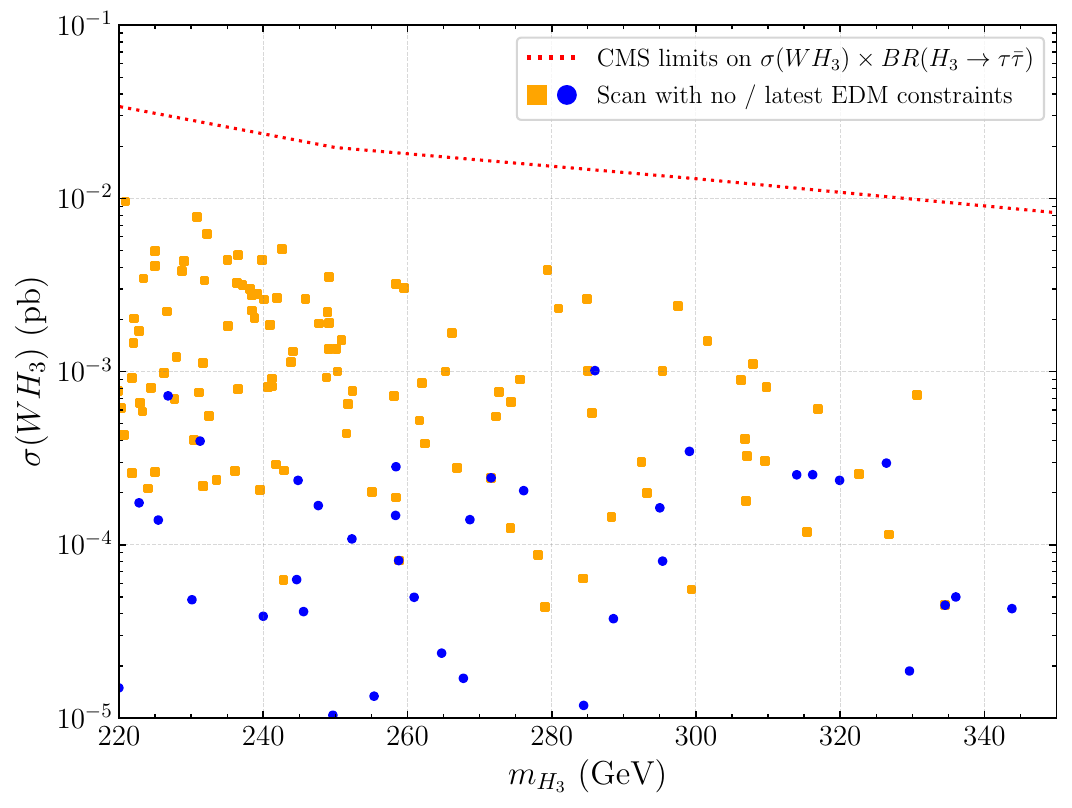} 

\vspace{-3mm}

\caption{\em Values of the $p p \to W H_3$ production cross section as a function of $m_{{H_3}}$ for the scan points potentially within reach of HL-LHC in Fig.~\ref{fig:GGFH3ZH1}. Also shown is the current CMS 95\% C.L. bound for the $H_3 \to \tau \tau$ final state \cite{CMS:2024ulc}, for reference.}
\label{fig:AssociatedWH3_NEW}
\par\end{centering}
\end{figure}

To determine the possibility of establishing $H$-CPV via set \textit{(v)}, we further collect the $S_2$ scan points of Fig.~\ref{fig:GGFH3ZH1} that could be probed at the HL-LHC via $g g \to H_3 \to Z H_1$ (using the strongest sensitivity projection), and show in Fig.~\ref{fig:AssociatedWH3_NEW} the corresponding values of the 
$p p \to W H_3$ cross section for these points, which can be as large as 10 fb. Existing searches for this BSM signature exist in the $H_a \to \tau\tau$ final state~\cite{CMS:2024ulc} (already discussed in this work and whose limits we show in Fig.~\ref{fig:AssociatedWH3_NEW} for reference), yet the $H_a \to \tau\tau$ branching fraction for these scan points lies well-below $10^{-3}$, so that this final state would not allow to probe $H$-CPV in this context. Another possibility is the $H_a \to W W$ final state (with branching fractions that can exceed 10\% in our scan), for which a search by ATLAS~\cite{ATLAS:2022eap} using same-sign di-leptons has been performed. Yet, the model interpretations shown in the ATLAS analysis do not apply to our scenario, and a robust sensitivity assessment is not possible. In particular, the results of Ref.~\cite{ATLAS:2022eap} are interpreted in an effective field theory framework with higher-dimensional operators, where the limits provided depend on anomalous couplings that vanish in our model. We leave such a potential study for the future.

\begin{figure}[t]
\begin{centering}
\includegraphics[width=0.49\textwidth]{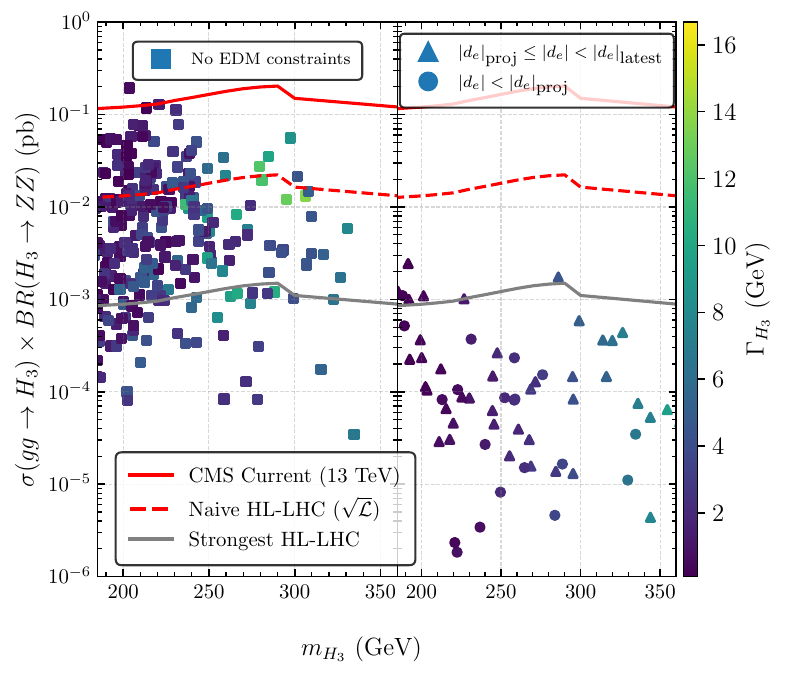}

\vspace{-3mm}

\caption{\em ggF-mediated $p p \to H_3 \rightarrow Z Z$ cross sections for the $S_2$ scan points above the strongest HL-LHC sensitivity extrapolation in both  Fig.~\ref{fig:AssociatedW} and Fig.~\ref{fig:GGFH3ZH1} (and satisfying $m_{H_3} > 2 \, m_Z$), with the total width $\Gamma_{H_3}$ for each point depicted in colour. Left (right) panels correspond to disregarding (including) the latest EDM constraints (triangles) and their projections $|d_e|_{\textrm{proj}}=1.0 \times 10^{-30} \, e \cdot \textrm{cm}$ (circles), see text for details. Also shown are the current 95\% C.L. limits on the $p p \to H_3 \rightarrow Z Z$ signature (solid-red line) from CMS~\cite{CMS:2018amk} -- fixing $\Gamma_{H_3} = 10$ GeV (the experimental limits depend very weakly on the width as long as the resonance is narrow) -- and the naive ($\sqrt{\mathcal{L}}$; dashed-red line) and strongest (solid-grey line) HL-LHC sensitivity extrapolations. 
}
\label{fig:GGFH3ZZ}
\par\end{centering}
\end{figure}

For the C2HDM scan points of Fig.~\ref{fig:GGFH3ZH1} above the HL-LHC (strongest) sensitivity, $H$-CPV through set \textit{(vi)} would require these scan points to also be visible via the processes $p p \to H_3 \to  V V$ and/or $ p p \to H_3 \to H_1 H_1$. We find that the most promising search in such case corresponds to $p p \to H_3 \to Z Z$, and show in Fig.~\ref{fig:GGFH3ZZ} the cross section values $\sigma(p p \to H_3) \times {\rm BR}(H_3 \to Z Z)$ for these points as a function of $m_{H_3}$ (for $m_{H_3} > 2\, m_Z$),  respectively disregarding/including the latest EDM constraints and their projections $|d_e|_{\textrm{proj}}$ in the left/right panels. We depict the current $\sqrt{s} = 13$ TeV limits from CMS~\cite{CMS:2018amk,b2gtwiki} (for a fixed total width $\Gamma_{H_3} = 10$ GeV; the limits depend weakly on the specific value of the width), together with the two HL-LHC sensitivity extrapolations used throughout this work (naive $\sqrt{\mathcal{L}}$ and strongest). Fig.~\ref{fig:GGFH3ZZ} highlights that, for the parameter space in which $m_{H_b} < m_h = 125$ GeV,  
set \textit{(vi)} -- through the combination of $p p \to W H_b$, $p p \to H_a \to Z H_b$ and $p p \to H_a \to Z Z$ signatures -- does constitute a viable avenue to probe $H$-CPV in the C2HDM, if EDM bounds are not considered at face value. We considered $|d_e|_{\textrm{proj}}=1.0 \times 10^{-30} \, e \cdot \textrm{cm}$, since the next-generation ACME III experiment is expected to improve the statistical sensitivity to the electron EDM by at least an order of magnitude, and potentially by a factor of 30~\cite{2022PhRvA.106b2808A, Wu:2022yqp}, compared to the current best limit from ACME II~\cite{ACME:2018yjb}. If we were to take $|d_e|_{\textrm{proj}}=0.3 \times 10^{-30} \, e \cdot \textrm{cm}$, we would be left with no points above the strongest HL-LHC line depicted in Fig.~\ref{fig:GGFH3ZZ}.

\begin{figure}[!h]
\begin{centering}
\includegraphics[width=0.49\textwidth]{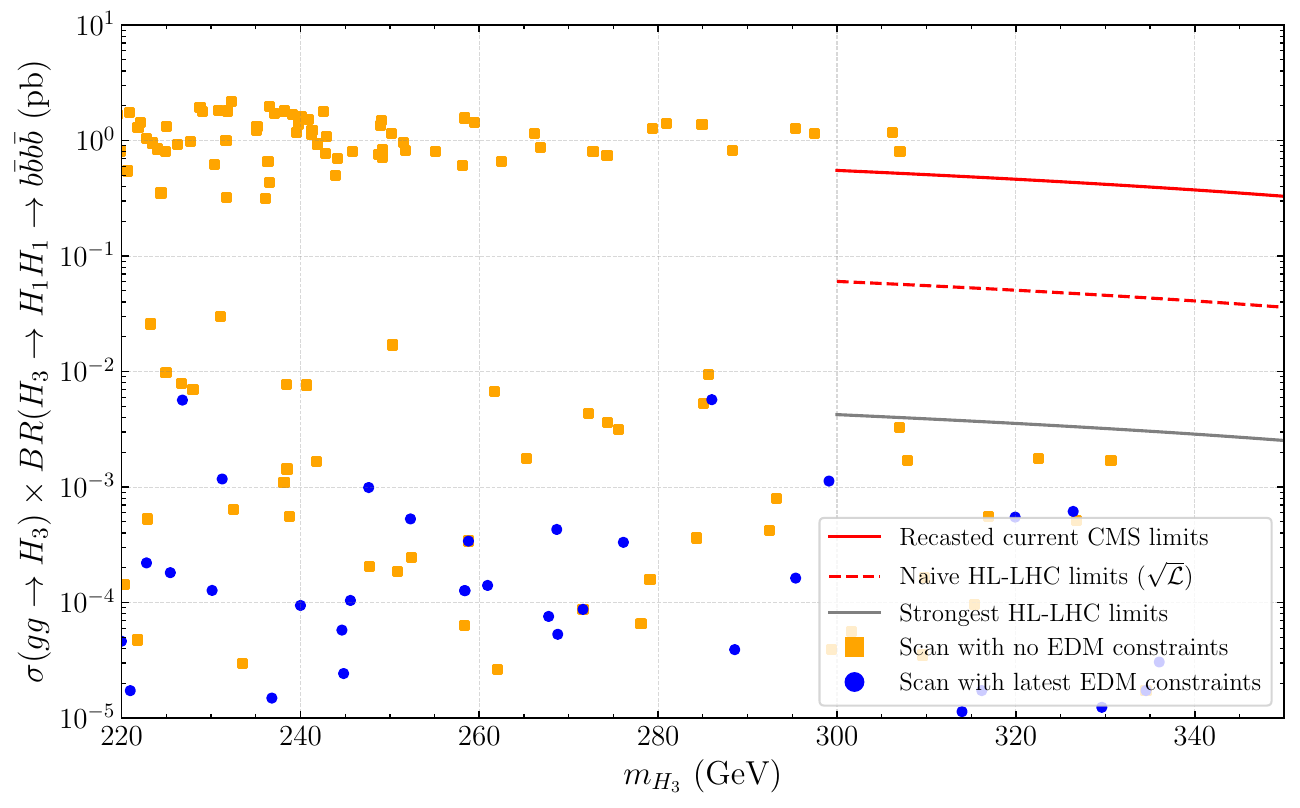}

\vspace{-3mm}

\caption{\em ggF-mediated $p p \to H_3 \rightarrow H_1H_1 \rightarrow b \bar{b} b \bar{b}$ cross sections for the $S_2$ scan points within HL-LHC reach from Fig.~\ref{fig:GGFH3ZH1}. Blue (Orange) points correspond to the inclusion (not inclusion) of the latest EDM constraints. Also shown are the recast $p p \to H_3 \rightarrow H_1H_1 \rightarrow b \bar{b} b \bar{b}$ 95\% C.L. limits from~\cite{Barducci:2019xkq} for 35.9 fb$^{-1}$ (solid-red line) and the naive ($\sqrt{\mathcal{L}}$; dashed-red line) and strongest (solid-grey line) HL-LHC sensitivity extrapolations, all for a benchmark value $m_{H_1}=100$ GeV.
}
\label{fig:GGFH3H1H1}
\par\end{centering}
\end{figure}

Besides $p p \to H_3 \to  Z Z$, another promising signature which may be used to complete the $H$-CPV set \textit{(vi)} is $p p \to H_3 \to H_1 H_1 \to b\bar{b} b\bar{b}$. The $p p \to H_3 \to H_1 H_1 \to b\bar{b} b\bar{b}$ cross section for the $S_2$ scan points of Fig.~\ref{fig:GGFH3ZH1} potentially within HL-LHC reach are then depicted in Fig.~\ref{fig:GGFH3H1H1}, exceeding in some cases $1$ pb. While there is no existing LHC experimental search for such a signature, a sensitivity study has been performed in Ref.~\cite{Barducci:2019xkq} through the recasting of CMS resonant di-Higgs searches in the $p p \to H_3 \to  h h \to b\bar{b} b\bar{b}$ final state with $35.9$ fb$^{-1}$ of integrated luminosity~\cite{CMS:2018qmt}. These CMS recast sensitivities are depicted in Fig.~\ref{fig:GGFH3H1H1} for a benchmark value $m_{H_1} = 100$ GeV (solid-red line), together with both the naive ($\sqrt{\mathcal{L}}$; dashed-red line) and strongest (solid-grey line) HL-LHC sensitivity extrapolations.
The recast sensitivities from Ref.~\cite{Barducci:2019xkq} only apply to $m_{H_3} > 300$ GeV, yet a potential ATLAS/CMS search would surely be able to probe lower masses, providing a complementary way to test $H$-CPV within set \textit{(vi)}. 

\begin{figure}[h]
\begin{centering}
\includegraphics[width=0.49\textwidth]{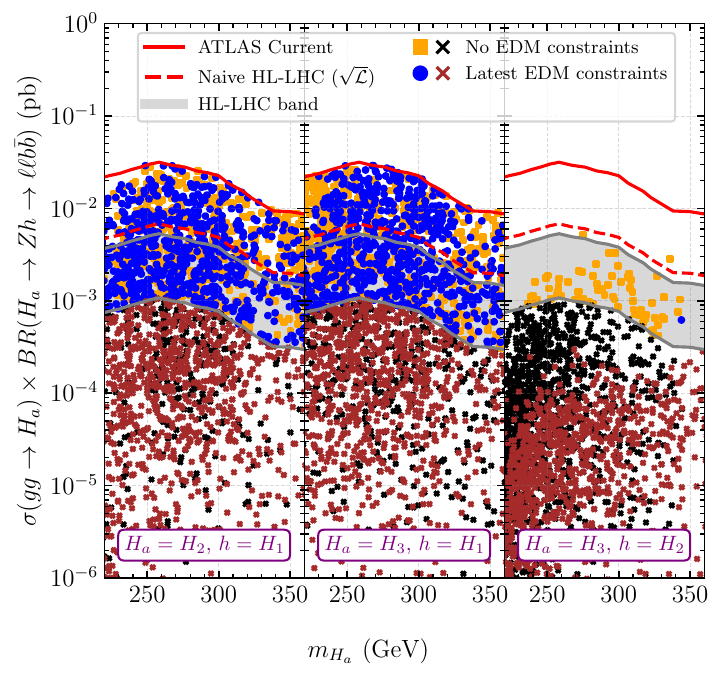} 

\vspace{-3mm}

\caption{\em ggF production $p p \to H_a \to Z h \to \ell \ell \, b \bar{b}$ cross sections at $\sqrt{s}=14$ TeV for scans $S_1$ (left and center panels) and $S_2$ (right panel), as a function of $m_{H_a}$. We show the current (solid red) and projected HL-LHC (dashed-red, grey band) sensitivity lines of this search, see text for details. Orange squares (black crosses) correspond to HL-LHC observable (unobservable) C2HDM scan points when no EDM constraints are applied, and blue circles (red crosses) when EDM bounds are strictly imposed.}
\label{fig:ggFh3_h3toZZ}
\par\end{centering}
\end{figure}

\begin{figure}[t]
\begin{centering}
\includegraphics[width=0.49\textwidth]{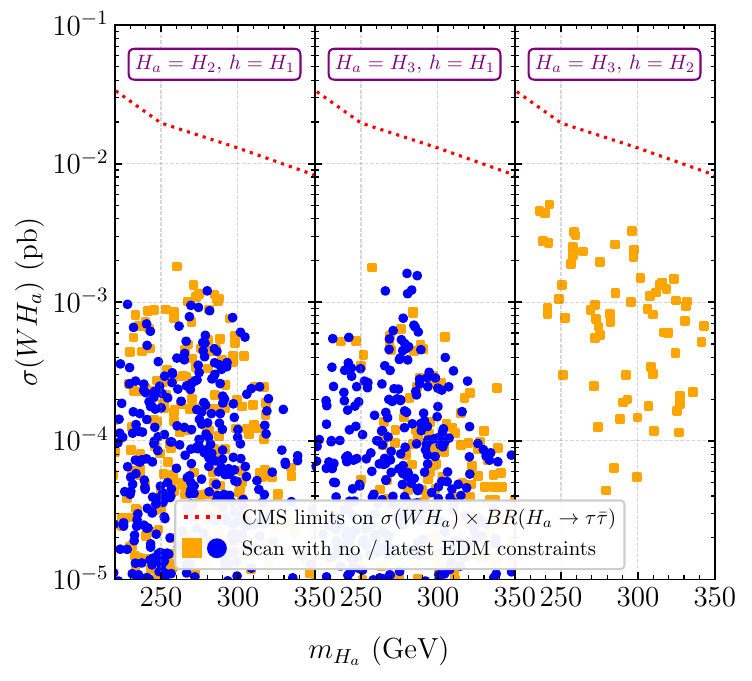}

\vspace{-3mm}

\caption{\em Values of the $p p \to W H_a$ production cross section as a function of $m_{{H_a}}$ for the observable scan points in Fig.~\ref{fig:ggFh3_h3toZZ}. Also shown is the current CMS 95\% C.L. bound for the $H_a \to \tau \tau$ final state \cite{CMS:2024ulc}, for reference. 
}
\label{fig:AssociatedWH3_casevii}
\par\end{centering}
\end{figure}

Finally, we discuss $H$-CPV via set \textit{(vii)}, which is independent of the results of Fig.~\ref{fig:AssociatedW}. We show the ggF $p p \to H_a \to Z h \to \ell \ell \, b \bar{b}$ cross sections in Fig.~\ref{fig:ggFh3_h3toZZ} for both $S_1$ (left, centre) and $S_2$ (right) scans (with either $H_a = H_3$ or $H_a = H_2$ depending on the scan). We also depict the present strongest LHC limits from the ATLAS search~\cite{ATLAS:2022enb} (solid-red line), as well as the HL-LHC naive ($\sqrt{\mathcal{L}}$; dashed-red line) sensitivity extrapolation from the present ATLAS search and the improved  HL-LHC (solid-grey band) sensitivity extrapolation discussed in App.~B. For scan $S_1$ the C2HDM $p p \to H_3 \to Z h \to \ell \ell \, b \bar{b}$ cross sections can saturate the current ATLAS bounds both including (blue points) and disregarding (orange points) the latest EDM constraints. For scan $S_2$ the cross sections are significantly lower, yet there are many points within potential reach of the HL-LHC if EDM constraints are not taken into account. For the $S_1$ and $S_2$ scan points above the strongest HL-LHC sensitivity extrapolation in Fig.~\ref{fig:ggFh3_h3toZZ}, we then show in Fig.~\ref{fig:AssociatedWH3_casevii} the resulting cross sections for $p p \to W H_a$. These cross sections can reach $\sim 10$ fb, and -- as already emphasized in the discussion around Fig.~\ref{fig:AssociatedWH3_NEW} -- the possibility of probing $W$-AP of $H_a$ at the HL-LHC in the $H_a \to W W$ decay mode 
seems promising. The recent search by ATLAS~\cite{ATLAS:2022eap} provides an important step in this direction, but the model interpretations do not encompass our scenario. The possibility to establish $H$-CPV via set \textit{(vii)} then remains an open question. 

\begin{figure}[h]
\begin{centering}
\includegraphics[width=0.45\textwidth]{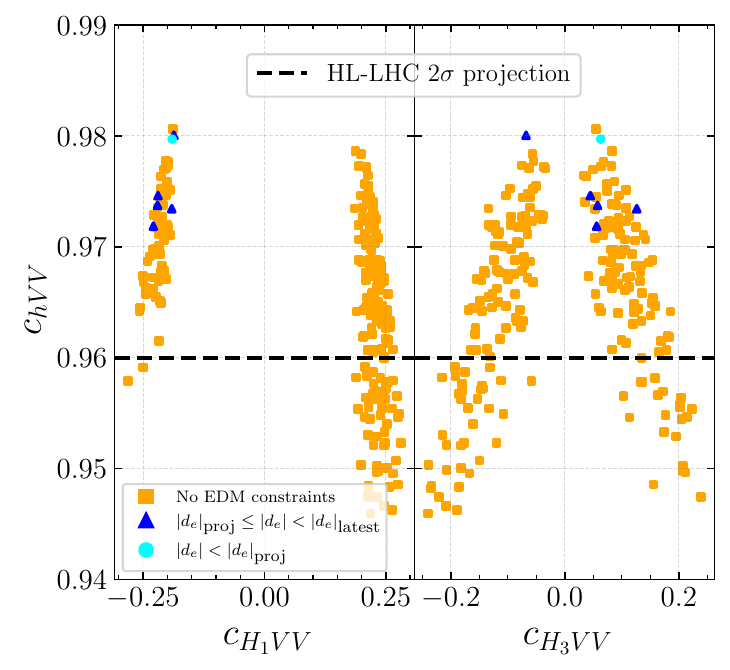}

\vspace{-3mm}

\caption{\em Values of the Higgs-gauge couplings $c_{H_1VV}$, $c_{hVV}$ and $c_{H_3VV}$ for the scan points passing the constraints of set \textit{(vi)} (for scan $S_2$). The dashed-black line showcases the CMS HL-LHC $2\sigma$ sensitivity projection on $c_{hVV}$ from Ref.~\cite{CMS:2022dwd}. Points without EDM constraints are shown as orange squares, those satisfying the latest EDM constraints as blue triangles, and those fulfilling the projected limit $|d_e|{\textrm{proj}}$ as cyan circles. The projected eEDM limit is $|d_e|_{\textrm{proj}} = 1 \times 10^{-30} \, e \cdot \textrm{cm}$, the same value used in Fig.~\ref{fig:GGFH3ZZ}.
}
\label{fig:CMS_chvv_chff}
\par\end{centering}
\end{figure}

Before concluding, we stress that, while all our $S_1$ and $S_2$ benchmark points satisfy present constraints from 125 GeV Higgs signal strength measurements on the properties of $h$, it is foreseeable that the future sensitivity on 125 GeV Higgs couplings of the HL-LHC will probe a significant fraction of the scan points for which $H$-CPV can be established via our various sets \textit{(i)}--\textit{(vii)}. Focusing for simplicity only on set \textit{(vi)}, which has been shown to be the most promising avenue analyzed in this work to test $H$-CPV within the C2HDM with a softly-broken $\mathbb{Z}_2$ symmetry, we show in Fig.~\ref{fig:CMS_chvv_chff} the value of $c_{hVV}$ for the scan points for which $H$-CPV can potentially be established at the HL-LHC (all belonging to scan $S_2$), together with their correlation with the values of $c_{H_1 VV}$ and $c_{H_3VV}$. The HL-LHC projection on the sensitivity to $c_{hVV}$ as obtained from Ref.~\cite{CMS:2022dwd} is also depicted (dashed-black line), highlighting that a sizeable fraction of the C2HDM scan points featuring observable $H$-CPV through the strategy investigated in this work, would however not lead to deviations in 125 GeV Higgs boson couplings measurable at HL-LHC.  

We end this section with two important remarks on
higher order processes. First of all, in a CP-conserving 2HDM, the decay $A \to ZZ$ is a fermionic loop-induced process and therefore very small. In this work we have been discussing a CP-violating 2HDM. However, if the CP-
violating couplings were so small that they would become comparable with loop-induced processes, the particle CP-numbers would have to be investigated in more
detail~\cite{Arhrib:2018pdi}. Yet, the values of $c_{H_1 VV}$ and $c_{H_3VV}$ shown in Fig.~\ref{fig:CMS_chvv_chff} suggest that this would not be an issue for our scan points with observable $H$-CPV at the HL-LHC. As a second comment, higher order corrections can modify the branching ratios
of the Higgs decays and in particular the ones to other Higgses~\cite{Krause:2016xku,Braathen:2019pxr}, although a dedicated analysis is beyond the scope of this paper.

\vspace{1mm}


\noindent \textbf{IV. Conclusions and outlook.}~In a scenario where CP-violation has its origin in a P-even, C-violating sector, three independent processes are needed to reach a contradiction on the CP-numbers of the scalars involved. It follows that the identification of a genuine CPV effect can only be made explicit by checking off a combination of processes. We have identified several combinations of such  processes where either the three neutral scalars from a C2HDM are present, or just two scalars together with the fact that the 125 GeV Higgs boson has been experimentally shown to contain a CP-even admixture. 
All these multiple signal strategies, listed in the sets \textit{(i)}--\textit{(vii)} of our manuscript, include VBF or $V$-associated production of BSM Higgs bosons as a core process, highlighting the potential key role of these processes in probing CP-violation within an extended Higgs sector. In particular, $W$-AP of BSM Higgs bosons is a key channel for probing $H$-CPV at the HL-LHC, which we believe is currently underexplored by the ATLAS and CMS collaborations. After taking into account the present constraints coming from collider searches, we have investigated all the proposed sets to understand if they could give us a hint of a CPV model at the HL stage of the LHC. We have concluded that the set \textit{(vi)} is the most promising and that there are good chances to find CPV in the Type I version of the C2HDM. We note that the analysis can be performed to not only any Yukawa types but also to more general classes of models where new sources of CPV have their origin in the scalar sector.

It is possible that to attain a stronger probe of such a CPV scenario, one would probably need the cleaner environment of a multi-TeV lepton collider. This in turn would provide decisive information about the nature of the extended scalar sector.

\vspace{3mm}
\begin{center}
\textbf{Acknowledgements} 
\end{center}
\vspace{-1mm}
\begin{acknowledgements}

Feynman diagrams were drawn using {\sc TikZ-Feynman}~\cite{Ellis:2016jkw}.~The authors thank Mia Tosi and Jorge Fern\'andez de Troc\'oniz for useful communications. A.L.O. thanks Marta F. Zamoro and Samuel Rosende Herrero for useful discussions. R.C. thanks Daniel Neascu for help in the implementation of the C2HDM. The authors thank Arturo de Giorgi for participation in early discussions of the project. ~A.L.O., L.M. and J.M.N. acknowledge partial financial support by the Spanish Research
Agency (Agencia Estatal de Investigaci\'on) through the grant IFT Centro de Excelencia Severo Ochoa CEX2020-001007-S and the grants PRE2022-105383, PID2022-137127NB-I00 and PGC2018-096646-A-I00 funded by
MCIN/AEI/10.13039/501100011033/FEDER, UE, and by the European Union’s Horizon
2020 research and innovation programme under the Marie Sklodowska-Curie grant agreement No 860881-HIDDeN and 101086085-ASYMMETRY. The work of J.M.N. is supported by the Ram\'on y Cajal Fellowship contract RYC-2017-22986. R.S. and R.C. are are partially supported by the Portuguese Foundation for Science and Technology (FCT) under CFTC: UIDB/00618/2020, 
UIDP/00618/2020,
and through the PRR (Recovery and Resilience
Plan), within the scope of the investment “RE-C06-i06 - Science Plus Capacity Building”, measure “RE-C06-i06.m02 - Reinforcement of financing for International Partnerships in Science,
Technology and Innovation of the PRR”, under the project with the reference 2024.03328.CERN. R.C. is additionally supported by FCT with a PhD Grant No.~2020.08221.BD. The authors acknowledge support from the COMETA COST Action CA22130.

\end{acknowledgements}

\section{Appendix A: Details on the C2HDM}
\label{AppB}

\subsection{The model}

The complex Two-Higgs doublet model (C2HDM) is one of the simplest scalar sector extensions of the SM to include new sources of CP-violation and has been the subject of many studies, see for instance Refs.~\cite{Khater:2003wq, ElKaffas:2007rq, Grzadkowski:2009iz, Arhrib:2010ju, Barroso:2012wz,Abe:2013qla, Inoue:2014nva, Cheung:2014oaa, Fontes:2014xva,Dorsch:2016nrg,Grober:2017gut,Fontes:2015xva, Fontes:2015mea,
Chen:2015gaa, Chen:2017com, Muhlleitner:2017dkd, Fontes:2017zfn, Basler:2017uxn, Aoki:2018zgq, Basler:2019iuu, Wang:2019pet, Boto:2020wyf, Cheung:2020ugr, Altmannshofer:2020shb, Fontes:2021iue, Basler:2021kgq, Frank:2021pkc, Abouabid:2021yvw,
Fontes:2022izp,deGiorgi:2023wjh,Azevedo:2023zkg,Goncalves:2023svb, Biekotter:2024ykp}.

The scalar potential of the C2HDM, given in Eq.~\eqref{eq:2HDM}, contains six real parameters, $m_{11}^2$, $m_{22}^2$, $\lambda_1$, $\lambda_2$, $\lambda_3$ and $\lambda_4$, as well as two complex parameters, $\lambda_5$ and $m_{12}^2$. Due to the hermiticity of the potential, only these two parameters can be complex. 

The Higgs potential is explicitly CP-violating since no choice of basis exists in which all the Higgs potential parameters are simultaneously real. Its properties can be studied in a charge-conserving vacuum by expanding the doublets around the vacuum expectation values (VEVs) $v_1$ and $v_2$ as
\begin{equation}
\begin{array}{c c c c c c}
    \Phi_1 = \begin{pmatrix}
        \phi_1^{+} \\
        \frac{v_1 + \rho_1 + i \eta_1}{\sqrt{2}}
    \end{pmatrix}
    & , &
  \Phi_2 = \begin{pmatrix}
        \phi_2^{+} \\
        \frac{v_2 + \rho_2 + i \eta_2}{\sqrt{2}} 
    \end{pmatrix}
    & , &
\end{array}
\end{equation}
where $\phi_i^+$ denotes the charged and $\rho_i$, $\eta_i$ the neutral scalar fields of the respective ($i=1,2$) doublets. Employing the conventional definition
\begin{equation}
    \lambda_{345} \equiv \lambda_3 + \lambda_5 + \Re(\lambda_5) \, ,
\end{equation}
the potential is minimized for
\begin{align}
    m_{11}^2v_1 + \frac{\lambda_1}{2}v_1^3 + \frac{\lambda_{345}}{2}v_1v_2^2 &= \Re(m_{12}^2) \, v_2 \\
    m_{22}^2v_2 + \frac{\lambda_2}{2}v_2^3 + \frac{\lambda_{345}}{2}v_1^2v_2 &= \Re(m_{12}^2) \, v_1 \\
    2 \, \Im(m_{12}^2) &= v_1 v_2 \, \Im(\lambda_5) \, .
\end{align}
The last condition is the reason why, despite the two complex phases, CP violation in this model is determined by \textit{one} independent parameter. As long as ${\rm Arg} [\lambda_5] \neq 2 {\rm Arg} [m_{12}^2]$, the two phases cannot be removed simultaneously. However, if ${\rm Arg} [\lambda_5] = 2 {\rm Arg} [m_{12}^2]$ and the VEVs of $\Phi_{1}$ and $\Phi_{2}$ are real, we are in the CP-conserving limit of the model. 

It is convenient to visualize this model in the Higgs basis, wherein the SM VEV and the Goldstone bosons are located in one of the doublets. To this aim, one rotates the original the doublets as
\begin{equation}
    \begin{pmatrix}
        \mathcal{H}_1 \\
        \mathcal{H}_2
    \end{pmatrix} =
    R^T_H \begin{pmatrix}
        \Phi_1 \\
        \Phi_2
    \end{pmatrix}
    \equiv
    \begin{pmatrix}
        c_{\beta} & s_{\beta} \\
        -s_{\beta} & c_{\beta}
    \end{pmatrix}
    \begin{pmatrix}
        \Phi_1 \\
        \Phi_2
    \end{pmatrix} \,.
\end{equation}
The rotation angle $\beta$ measures the ratio of the VEVs as
\begin{equation}
    \tan \beta \equiv \frac{v_2}{v_1} \, .
\end{equation}
In this basis, the doublets are reexpressed as
\begin{equation}
\begin{array}{c c c c c c}
    \mathcal{H}_1 = \begin{pmatrix}
        G^{\pm} \\
        \frac{1}{\sqrt{2}} (v + H^0 + i G^0)
    \end{pmatrix}
    & , &
    \mathcal{H}_2 = \begin{pmatrix}
        H^{\pm} \\
        \frac{1}{\sqrt{2}} (R_2 + i I_2)
    \end{pmatrix}
    & , &
\end{array}
\end{equation}
where we have defined the total (SM) VEV as
\begin{equation}
    v = \sqrt{v_1^2 + v_2^2} \, .
\end{equation}
For a study of the spectrum of neutral states it is convenient to work in the mass basis. Denoting the mass eigenstates as $H_i$ for $i\in \{1,2,3\}$, this is achieved by appropriate rotations
\begin{equation}
    \begin{pmatrix}
        H_1 \\
        H_2 \\
        H_3
    \end{pmatrix}
    = R \, \tilde{R}_H
    \begin{pmatrix}
        H^0 \\
        R_2 \\
        I_2
    \end{pmatrix}
    = R
    \begin{pmatrix}
        \rho_1 \\
        \rho_2 \\
        \rho_3
    \end{pmatrix}
    \, ,
\end{equation}
where we fix $\rho_3 = I_2$ and have embedded the Higgs-to-$\Phi$-basis rotation matrix $R_H$ into
\begin{equation}
    \tilde{R}_H = \begin{pmatrix}
        R_H & 0\\
        0 & 1
    \end{pmatrix} \, .
\end{equation}
Given the Hessian of the potential with respect to the neutral states in the $\Phi$-basis
\begin{equation}
    (\mathcal{M}^2)_{ij} = \left< \frac{\partial^2 V}{\partial \rho_i \partial \rho_j}\right> \, ,
\end{equation}
the masses of the neutral scalar eigenstates are obtained as
\begin{equation}
    R\mathcal{M}^2R^T = \diag(m_{H_1}^2, m_{H_2}^2, m_{H_3}^2) \, ,
\end{equation}
where a hierarchy $m_{H_1} \leq m_{H_2} \leq m_{H_3}$ is assumed. 

Using the standard trigonometric notation $(s_i, c_i) = (\sin \alpha_i, \cos \alpha_i)$ and the ranges
\begin{equation}
    -\pi/2 < \alpha_i \leq \pi/2 \, ,
\end{equation}
for $i=1,2,3$, the rotation matrix $R$ can be parametrized as \cite{Accomando:2006asw,Khater:2003wq,ElKaffas:2007rq,Fontes:2017zfn}
\begin{equation}
\label{rotation_matrix}
    \begin{pmatrix}
        c_1 c_2 & s_1 c_2 & s_2 \\
        -(c_1s_2s_3 + s_1c_3) & c_1c_3 - s_1s_2s_3 & c_2s_3 \\
        -c_1s_2c_3 + s_1s_3 & -(c_1s_3 + s_1s_2c_3) & c_2c_3
    \end{pmatrix} \, .
\end{equation}

After fixing all parameter relations, the potential possesses nine free parameters that can be traded for nine physical input parameters, namely $m_{H_1}$, $m_{H_2}$, $m_{H^{\pm}}$, $\Re(m_{12}^2)$, $\alpha_1$, $\alpha_2$, $\alpha_3$, $v$ and $\tan \beta$. In particular, the above choices imply the following relation for the mass of the remaining neutral eigenstate,
\begin{equation}
    m_{H_3}^2 = \frac{m_{H_1}^2 \, R_{13}(R_{12} t_\beta - R_{11}) + m_{H_2}^2 \, R_{23}(R_{22} t_\beta - R_{21})}{R_{33}(R_{31} - R_{32} t_\beta)}
\end{equation}
with the restriction $m_{H_3} > m_{H_2}$.

The Higgs couplings to the massive gauge bosons $V=W,Z$ are given by
\begin{equation}
    i \, g_{\mu\nu} \, c_{H_i VV} \, g_{h VV} \;, \label{eq:gaugecoupdef}
\end{equation}
where $g_{hVV}$ denotes the corresponding SM Higgs coupling,
given by
\begin{equation}
g_{hVV} = \left\{\begin{array}{c} g M_W \qquad \qquad \quad V=W
\\ g M_Z /\cos\theta_W \qquad V=Z \end{array} \right. \qquad
\end{equation}
where $g$ is the $SU(2)$ gauge coupling and $\theta_W$ is the Weinberg angle.
The effective couplings can be written as
\begin{equation}
c_{H_i VV} = c_\beta R_{i1} + s_\beta R_{i2} \;. \label{eq:c2dhmgaugecoup}
\end{equation}

The Yukawa sector is built by extending the $\mathbb{Z}_2$ symmetry to the fermion
fields such that flavour changing neutral currents (FCNC) are absent at
tree-level~\cite{Glashow:1976nt,Paschos:1976ay}.
There are four possible $\mathbb{Z}_2$ charge assignments and therefore four
different types of 2HDMs described in Tab.~\ref{tab:types}.

\begin{table}
\begin{center}
\begin{tabular}{rccc} \hline
& $u$-type & $d$-type & leptons \\ \hline
Type I & $\Phi_2$ & $\Phi_2$ & $\Phi_2$ \\
Type II & $\Phi_2$ & $\Phi_1$ & $\Phi_1$ \\
Lepton-Specific & $\Phi_2$ & $\Phi_2$ & $\Phi_1$ \\
Flipped & $\Phi_2$ & $\Phi_1$ & $\Phi_2$ \\ \hline
\end{tabular}
\caption{\em The four Yukawa types of the softly broken $\mathbb{Z}_2$-symmetric 2HDM. \label{tab:types}}
\end{center}
\end{table}

The Yukawa Lagrangian has the form
\begin{equation}
{\cal L}_Y = - \sum_{i=1}^3 \frac{m_f}{v} \bar{\psi}_f \left[ c^e_{H_i
  ff} + i c^o_{H_i ff} \gamma_5 \right] \psi_f H_i \;, \label{eq:yuklag}
\end{equation}
where $\psi_f$ denotes the fermion fields with mass $m_f$. The
coefficients of the CP-even and of the CP-odd part of the Yukawa
coupling, $c^e_{H_i ff}$ and $c^o_{H_i ff}$, respectively, are presented in Tab.~\ref{tab:yukcoup}. A longer list of the interactions between the Higgs bosons and the remaining SM fields can be found in Ref.~\cite{Fontes:2017zfn}.

\begin{table}
\centering 
\footnotesize
\begin{tabular}{rccc} \hline
& $u$-type & $d$-type & leptons \\ \hline
Type I & $\frac{R_{i2}}{s_\beta} - i \frac{R_{i3}}{t_\beta} \gamma_5$
& $\frac{R_{i2}}{s_\beta} + i \frac{R_{i3}}{t_\beta} \gamma_5$ &
$\frac{R_{i2}}{s_\beta} + i \frac{R_{i3}}{t_\beta} \gamma_5$ \\
Type II & $\frac{R_{i2}}{s_\beta} - i \frac{R_{i3}}{t_\beta} \gamma_5$
& $\frac{R_{i1}}{c_\beta} - i t_\beta R_{i3} \gamma_5$ &
$\frac{R_{i1}}{c_\beta} - i t_\beta R_{i3} \gamma_5$ \\
Lepton-Specific & $\frac{R_{i2}}{s_\beta} - i \frac{R_{i3}}{t_\beta} \gamma_5$
& $\frac{R_{i2}}{s_\beta} + i \frac{R_{i3}}{t_\beta} \gamma_5$ &
$\frac{R_{i1}}{c_\beta} - i t_\beta R_{i3} \gamma_5$ \\
Flipped & $\frac{R_{i2}}{s_\beta} - i \frac{R_{i3}}{t_\beta} \gamma_5$
& $\frac{R_{i1}}{c_\beta} - i t_\beta R_{i3} \gamma_5$ &
$\frac{R_{i2}}{s_\beta} + i \frac{R_{i3}}{t_\beta} \gamma_5$ \\ \hline
\end{tabular}
\caption{\em Yukawa couplings of the Higgs
  bosons $H_i$ in the C2HDM, divided by the corresponding SM Higgs couplings. The expressions correspond to
  $[c^e_{H_i ff} +i c^o_{H_i ff} \gamma_5]$ from
  Eq.~(\ref{eq:yuklag}). \label{tab:yukcoup}}
\end{table}

\subsection{Theoretical and experimental constraints}

In this section we briefly summarize the main theoretical and experimental restrictions that we considered in the C2HDM scans (for more details about these constraints, see for instance Refs.~\cite{Fontes:2017zfn, Biekotter:2024ykp}). 
The C2HDM was implemented in
{\tt ScannerS}~\cite{Coimbra:2013qq,Muhlleitner:2020wwk}.
The most relevant theoretical and experimental bounds are either built
in the code or acessible via interfaces with other codes. We have imposed the following constraints on the scans that were performed:

\begin{itemize}
    \item[-] The theoretical bounds include boundedness from below and perturbative unitarity~\cite{Kanemura:1993hm, Akeroyd:2000wc,Ginzburg:2003fe}, as well as vacuum stability, which can be accomplished by forcing the minimum to be global~\cite{Ivanov:2015nea}.
    \item[-] The points generated are required to comply with the measured values of the oblique parameters $S$, $T$ and $U$~\cite{Branco:2011iw}, within $2\sigma$ of the experimental results from \cite{Baak:2014ora}.
    \item[-] The most constraining bounds on the charged sector of the C2HDM come from the measurements of $B \to X_s \gamma$
\cite{Deschamps:2009rh,Mahmoudi:2009zx,Hermann:2012fc,Misiak:2015xwa,Misiak:2017bgg,Misiak:2020vlo}, which translate into $2\sigma$ exclusion bounds on the $m_{H^\pm}-t_\beta$ plane. We follow \cite{Fontes:2017zfn} in using $80 \mbox{ GeV} \le m_{H^\pm}< 1 \mbox{ TeV}$ and $0.8 \le t_{\beta} \le 35$ for the Type-I model that was considered in this work. Notice, however, that for Type-II and Flipped models, the bounds are much more severe, forcing $m_{H^\pm} > 580 \mbox{ GeV}$ almost independently of $\tan \beta$. 
\item[-] The most recent limit of $4.1\times 10^{-30} \, \text{e.cm}$
measured at JILA~\cite{Roussy:2022cmp} for the eEDM is considered. This limit was not applied to all the points. In the text and in the figures we indicate when the eEDM limit was applied in our scans. 
\item[-] Compatibility with new searches for additional scalars and LHC constraints on the discovered Higgs boson at the LHC are checked via the interface to the code \texttt{HiggsTools}~\cite{Bahl:2022igd}. As mentioned in the main text, we apply an additional filtering to our scan points by requiring that they lie within the $(c_{hff},c_{hVV})$ $2\sigma$ ellipse in Ref.~\cite{CMS:2022dwd} from a combined Higgs signal strength fit with observed best-fit value $(c_{hff},c_{hVV}) = (0.906,1.014)$.
\end{itemize}

\section{Appendix B: HL-LHC sensitivities}
\label{AppA}

As outlined in the main text, the naive ``statistical'' $\sqrt{\mathcal{L}}$ extrapolation of the current experimental sensitivity of ATLAS and CMS analyses to the HL-LHC could turn out to be rather pessimistic in many cases, as discussed in detail in Ref.~\cite{Belvedere:2024wzg}. Such an extrapolation does not account for several ingredients that would presumably increase the future LHC sensitivity with respect to a $\sqrt{\mathcal{L}}$ projection: 
\begin{itemize}
\item[-] The possible inclusion of further decay modes in analyses already targeting a specific decay channel as $\mathcal{L}$ increases.
\item[-] The exploitation of kinematic regions (e.g.~high-$p_T$ configurations) that do not appreciably contribute to the present sensitivity, but would very much do for larger $\mathcal{L}$. 
\item[-] The development of new (e.g.~artificial-intelligence-based) analysis strategies. 
\item[-] The (obvious) increase in center-of-mass energy $\sqrt{s}$ from LHC Run II to HL-LHC, which will increase the sensitivity (albeit not in a dramatic way) of many BSM searches.
\end{itemize}

A robust estimate of the sensitivity increase at the HL-LHC with respect to the present ATLAS and CMS searches that we have analyzed lies beyond the scope of this work, and may not even be possible in some cases (e.g.~if new, machine-learning-based analysis tools were to be used in future searches). Yet, the sensitivity evolution of several ATLAS/CMS searches up to their current status offers a way to ascertain the possibility of improving over the $\sqrt{\mathcal{L}}$ extrapolation. Specifically, we parametrize the 
experimental sensitivity increase observed between an integrated luminosity $\simeq 36$ fb$^{-1}$ and $\simeq 140$ fb$^{-1}$ in different ATLAS and CMS SM searches performed for $\sqrt{s} = 13$ TeV LHC,\footnote{Avoiding the sensitivity comparison with $\sqrt{s} = 8$ TeV data allows to drop out of the discussion the sensitivity gain in going from $\sqrt{s} = 8$ TeV to $\sqrt{s} = 13$ TeV LHC c.o.m. energy.} 
as an $\mathcal{L}^x$ function of the luminosity increase $\mathcal{L} \simeq 140/36$ (with $x = 0.5$ corresponding to the ``statistical'' sensitivity increase). The searches we consider are: $p p \to t\bar{t} t\bar{t}$~\cite{CMS:2017ocm,CMS:2023ftu}, for which the sensitivity increase yields $x = 0.86$; $p p \to V h$ with $h\to b\bar{b}$~\cite{ATLAS:2017cen,ATLAS:2020fcp}, for which the sensitivity increase yields $x = 0.60$; EW vector boson scattering $p p \to Z Z + j j$, $ Z Z \to \ell\ell \ell'\ell'$ (with $\ell,\ell' = e,\mu$)~\cite{CMS:2017zmo,CMS:2020fqz}, for which the sensitivity increase yields $x = 0.58$; di-Higgs production $p p \to h h \to b\bar{b} \tau\tau$~\cite{ATLAS:2018uni,ATLAS:2024pov}, for which the sensitivity increase yields $x = 1.11$. We note that all these searches yield a sensitivity increase from $36$ fb$^{-1}$ to $140$ fb$^{-1}$ which exceeds the $\sqrt{\mathcal{L}}$ scaling. Then, as an ``optimistic'' projection of the sensitivity of the various searches we discuss in this work to the HL-LHC with 3000 fb$^{-1}$ of integrated luminosity, we use an $\mathcal{L}^x$ scaling, with $x \in [0.58, 1.11]$, rather than the naive $\sqrt{\mathcal{L}}$ scaling. This defines the ``improved'' HL-LHC extrapolated sensitivity band shown in grey in Fig.~\ref{fig:AssociatedW}.

\bibliography{CPVLib}

\end{document}